# Counterions in RNA structure: Structural bioinformatics analysis to identify the role of Mg2+ ions in base pair formation

**Swati Adhikari[1, a], Dhananjay Bhattacharyya[2, 3, b] and Parthajit Roy[1, c, *]**

[1] Department of Computer Science, The University of Burdwan, Golapbag, Purba Bardhaman, West Bengal, India, PIN: 713104

[2] Computational Science Division, Saha Institute of Nuclear Physics, 1/AF Bidhannagar, Kolkata, West Bengal, India, PIN: 700064 and

[3] Department of Biophysics, Molecular Biology and Bioinformatics, University of Calcutta, 92 A.P.C. Road, Kolkata, West Bengal, India, PIN: 700009

[a]Email: swatidhkr@gmail.com

[b]Email: bhattasinp@gmail.com, dhananjay.bhattacharyya.retd@saha.ac.in

[c]Email: roy.parthajit@gmail.com

*Corresponding Author

**Abstract**

**Context**

Contribution of metal ions on nucleic acids' structures and functions is undeniable. Among the available metal ions like Na+, K+, Ca2+, Mg2+ etc., the role that play the Mg2+ ion is very significant related to the stability of the structures of RNA and this is quite well studied. But it is not possible to grasp the entire functionality of Mg2+ ion in the structure of RNA. So, to have a better understanding of the Mg-RNA complexes, in the present study, we have investigated 1541 non-redundant crystal structures of RNA and generated reports for various statistics related to these Mg-RNA complexes by computing base pairs and Mg2+ binding statistics. In this study, it has also been reported whether the presence of Mg2+ ions can alter the stability of base pairs or not by computing and comparing the base pairs' stability. We noted that the Mg2+ ions do not affect the canonical base pair G:C W:WC while majority of the non-canonical base pair G:G W:HC, which is important also in DNA telomere structures, has Magnesium ion binding to O6 or N7 atoms of one of the Guanines.

**Methods**

To carry out the investigation, the RNA crystal structures are taken from the Nucleic Acid Database (NDB) Server. To compute base pairs, the MetBP software is used and this is followed by the application of the BPFIND algorithm. In the present study, all figures are created in PyMol.

## I. Introduction

Mg2+ ions have a pivotal role in the structure of nucleic acids. This ion is involved in various fundamental processes like replication of DNA, conversion of DNA into messenger RNA (mRNA), known as transcription and decoding of mRNA into proteins, known as translation [1]. Divalent cations like Mg2+ are also essential for stabilizing DNA and RNA structures [2-3]. Every and Russu had addressed this issue related to DNA structures in their research by a study of the DNA duplex and its interaction with Mg2+ ions [2]. The same concept related to RNA structures has been discussed by Misra and Draper [3]. Here the authors have shown the mechanism of stabilization of a particular folded form of RNA by the Mg2+ ions. Mukherjee and Bhattacharyya have shown that non electrostatic contribution plays a significant role while binding of Mg2+ ions with DNA [4]. As nucleic acids are polyanions, there is a requirement of positively charged counterions so that negatively charged phosphate groups be neutralized [5]. There is a competition between metal ions to bind with nucleic acids. As there exist more positive charges and higher hydration energy in Mg2+ ions than some other ions like Na+, K+, etc., Mg2+ ions win in this competition. Mg2+ ions thus assist in neutralizing the negative charge of the backbone. In interaction between Mg and DNA, Mg2+ ions interact with purine bases at the N7 region and pyrimidine bases at the N3 regions through coordinate bonds Mg-N7 and Mg-N3. Negatively charged O2 atoms of the phosphate group of nucleotide chains also interact with Mg2+ ions [5]. Such interactions help in stabilizing the higher order structures of DNA as well as stabilize the structure of the RNA and form RNA-protein complexes. This is important in the translation process.

To regulate the processes of transcription, translation etc. the RNA folds into higher order structures where Mg2+ ions interact with the ribose-phosphate backbone of RNA which is negatively charged and are responsible for neutralizing this negative charge of the backbone which helps in RNA folding in turn [6]. Halder et al. have discussed the importance of the Mg2+ ion binding mode in framing of the structure of RNA, its folding and function [7]. They have also observed how Mg2+ ions binding affects the geometry and stability of various non canonical base pairs of RNA structures [8]. During translation, the concentration of Mg2+ ions help in determining final protein products [9].

Mg2+ ions are also important in active sites of RNA polymerase which generally contains two such ions [10]. Here, one Mg2+ ion is required for deprotonation of the 3' OH group of the terminal nucleotide which allows the Mg2+ ion to interact with the phosphate bond. In this way the phosphodiester bond is formed and pyrophosphate is released. The second Mg2+ ion is required for stabilization of the transition state and it neutralizes the negative charge on the pyrophosphate group that has been released as a result of the earlier step.

Mg2+ ions are essential for regulating association and dissociation of subunits like nucleotides, bases, phosphate groups in mRNA. This ion act as a key player in stabilization of the RNA structures. Main components of nucleic acids are sugar, phosphate and nitrogenous bases. The common sugar and phosphate moieties form the negatively charged backbone while four different bases give rise to variability in terms of base sequence and structure-function as well. These bases mostly exist as paired with each other by following complementary base pairing rules. Five types of nitrogenous bases are found in DNA and RNA which are known as purines such as A (Adenine) & G (Guanine), and pyrimidines such as C (Cytosine), T (Thymine) & U (Uracil). Among these bases, A, G & C are present in DNA as well as in RNA and T & U are present only in DNA and in RNA respectively. Commonly observed base pairs, also known as canonical base pairs, are G:C (in DNA as well as in RNA), A:T (in DNA) and A:U (in RNA). Base pairs are stabilized by the number of hydrogen bonds (three or two) present between them. The bond through which one nitrogen base is attached with the Sugar-Phosphate group is known as the Glycosidic bond.

Each of the five bases has three hydrogen bonding edges through which two bases can interact [10]. These three edges are namely, Watson-Crick (W), Hoogsteen (H) & Sugar (S). In a base pairing, as per the position of the Glycosidic bonds, two orientations are available — cis orientation and trans orientation. In the cis orientation, two Glycosidic bonds are situated on the same side of the reference line whereas in the trans orientation, two Glycosidic bonds are situated on the opposite sides of the same. The Interacting edges of a purine and a pyrimidine are shown in Figures 1(a) and 1(b) respectively.

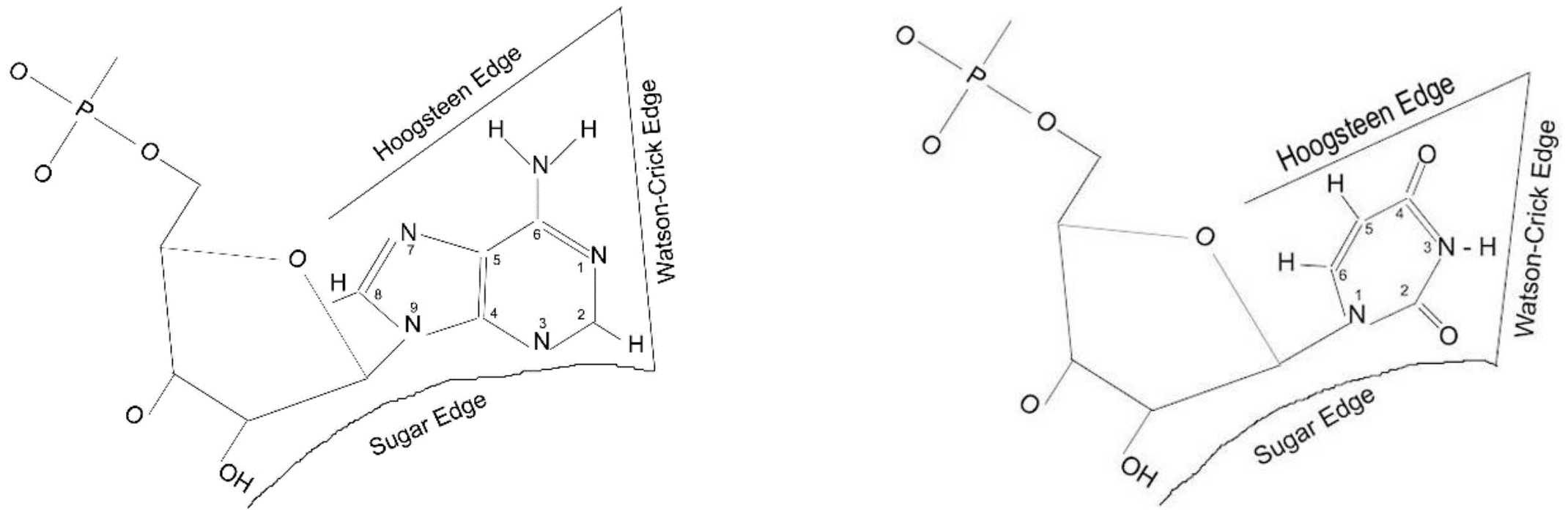

(a) Purine (b) Pyrimidine

Figure 1: Interacting edges

Along with the canonical base pairs, also known as Watson-Crick base pairs, several other base pairs (non-canonical / non-Watson-Crick) stabilized by multiple hydrogen bonds have been detected in the experimentally determined nucleic acid structures. According to hydrogen bonding edges and the position of Glycosidic bonds with respect to the hydrogen bonds in a base pairing geometry, twelve types of base pairing families are found [11]. These twelve types of base pairing families are namely, cis W:W, trans W:W, cis W:H, trans W:H, cis W:S, trans W:S, cis H:H, trans H:H, cis H:S, trans H:S, cis S:S, trans S:S.

Let us now draw our attention to Mg2+ ions and its interaction with RNA. According to Zheng and coworkers [12], the atoms of RNA which can coordinate with Mg2+ ions are categorized into four different types: (i) phosphate oxygen OP1/OP2, (ii) oxygens in ribose O2′/O4′ or it may be oxygens that are bridging phosphate and ribose O3′/O5′, (iii) oxygen of nucleobase and (iv) Nitrogen of nucleobase. In Mg-RNA interactions, both outer sphere and inner sphere modes are employed in binding of Mg2+ ions with RNA [13]. Outer sphere mode of binding (where Mg2+ binds to water oxygen and the water hydrogen atoms form hydrogen bonds with nucleic acids) is mainly seen in the major groove of A-DNA [13]. In this type of binding, the Mg2+ ion binds with negatively charged phosphate backbone to neutralize its charge wherein the preferred ligand atoms are phosphate oxygen atoms and atoms of bases which are electronegative in nature. Generally, hydrated Mg2+ ions form hydrogen bonds mostly with N7 & O6 atoms of guanine in GpG & GpU (where p indicates backbone phosphate group) sites and bind to the floor of the deep major groove of A-DNA [14]. The Hoogsteen edge of the guanine is the favorable location of Mg2+ ions to bind [13]. It happens mostly when guanine becomes part of the non-canonical base pair, although no magnesium binding is found in GpU or UpG in the crystal structure of the recognition area [15]. In an inner sphere binding, Mg2+ ion is directly linked with the RNA fragment. Figure 2 shows arrangement of atoms in two binding modes of Mg2+ ions with RNA in the structure 1f27.pdb [Chain ID: A, Residue Name: MG, Residue No.: 35]. This figure is created in PyMol [16].

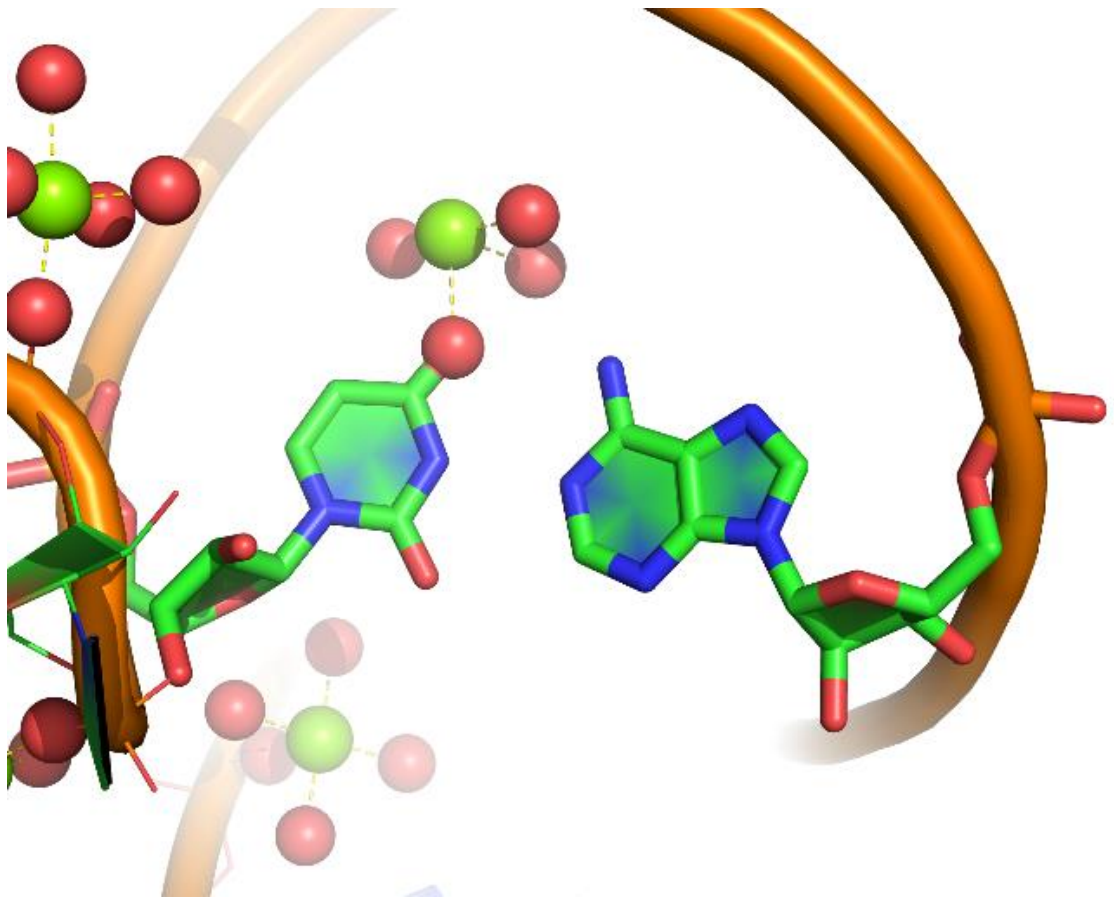

Figure 2: Different modes of Mg-RNA Binding – PDB ID: 1f27 [Chain ID: A, Residue Name: MG, Residue No.: 35]

Petrov et al. have discussed different conformations of water-mediated magnesium-guanine interactions [17]. Here, the authors have measured the interaction energy for eight complexes where tetra-hydrated, penta-hydrated and hexa-hydrated Mg2+ ions were primarily placed in the guanine plane near the N7 and O6 binding atoms. According to their results, the hexa-hydrated complex is considered as more stable than the others.

Thus, Mg-RNA interaction is quite well studied due to its immense importance. The available literature, however, does not give complete idea about the functionality of Mg2+ ions in the structure of RNA. This is possible due to several factors, such as unavailability of sufficient number of high-resolution structures, absence of proper computational tools to identify the base pairs and their contacts with ions, etc. In the present study, we shall investigate several crystal structures of RNA for having Mg-RNA complexes and generate reports for various statistics related to these Mg-RNA complexes by computing base pairs. It will also be reported whether the presence of Mg2+ ions is capable of altering the stability of base pairs or not by computing stability of the base pairs.

## II. Materials and Methods

We shall carry out our investigation on RNA crystal structures, 1541 structures in total, that have a resolution less than 3.0 Å. These structures are taken from the Nucleic Acid Database (NDB) Server [18], available as a non-redundant list (NRList-3.82) of RNA crystal structures. Among these 1541 structures, we shall only consider the structures in which Mg2+ ions are present and, in these structures, the presence of base pairs has been identified or detected.

To carry out the investigation, while identifying the base pairs, we have considered the following four types of base pairs:
(i) Normal polar hydrogen-bond donor-acceptor base pairs.
(ii) C-H- - - O/N interaction mediated base pairs.
(iii) Protonated base pairs.
(iv) Sugar-O2' mediated base pairs.

It should be noted that the direct hydrogen bonds between the bases are only considered and water-mediated base pairs are not identified. Furthermore, a base pair is identified only when two hydrogen bonds between the bases are possible. In this study, the base pairs involving modified bases are also considered.

The maximum limiting distance between two hydrogen bonding atoms is taken as 3.8 Å. Along with this, the following distance parameters are set for bond length between the Mg2+ and other atoms:
(i) Van der Waals radius of the Mg2+ is set to 2.22 Å
(ii) Maximum Mg-O (oxygen of nucleic acid) distance is set to 2.58 Å
(iii) The Mg-O (oxygen of $H_2O$ molecule) distance is set to 2.60 Å
(iv) The Mg-N distance is set to 2.70 Å

To compute base pairs, we have used the MetBP [19] software that follows the BPFIND algorithm [20]. Following statistics are obtained throughout our investigation and are reported in this study:
(i) If Mg2+ is present near any nucleotide moiety like base/sugar/phosphate of any base pair, then that is reported.
(ii) If Mg2+ interacts with any base which is not in any base pair, then that one is also reported separately.
(iii) The distance between the Mg2+ and the atom of a base pair with which the Mg2+ is bound are computed and reported.
(iv) The report regarding interaction of Mg2+ with base/sugar/phosphate and the atoms which are involved in such interaction are also generated.
(v) The report regarding the position of a base pair in the secondary structure of RNA like helix, coil, loop etc. with which Mg2+ has been interacted, is also generated.
(vi) The report is also generated for binding of Mg2+ with the atoms of any one edge among the three edges (W, H and S) of a base.
(vii) If any nucleic acid forms chelate with the Mg2+, then that is also reported.

Besides these, we have tried to show which types of bases are suitable for binding. All we know, the coordination number of Mg2+ is 6. But it is quite difficult to generate reports regarding all atoms in a crystal structure. So, we have done a thorough investigation on coordination numbers and in order to consider the coordination number, we have checked whether the Mg2+ has been interacted with the probable nucleic acids/proteins/waters/metals (Mg & others) and also checked whether there is any coordination as per such interaction.

We have calculated the stability of the base pair in order to see whether the presence of metal distorts the stability of the base pair or not. To calculate such stability, we have used the equation 1:

$$E = \sum_i (\Delta_i - 3.0)^2 - \sum_j (\varphi_j - \pi)^2 \qquad (1)$$

where $\Delta_i$ stands for the distance between two hydrogen bonding atoms and $\varphi_j$ stands for the pseudo angle involving suitably chosen precursor atoms and the two atoms involved in the hydrogen bond. Here, we have calculated the average stability in presence of metal and the average stability without metal. Instead of considering all types of base pairs at a time, specific base pair wise calculation has been done.

We have analyzed these data through our in-house programme.

## III. Results and Discussion

### III. 1 Interaction of Mg2+ ions with unpaired & paired residues

To carry out our investigation, first of all we shall present the general statistics regarding Mg-binding of nucleic acids related to bases that form base pairs and the bases that do not form base pairs.

Among the 1541 structures which are selected for our investigation, we found that 844 structures contain Mg2+ ions. The list of these 844 structures is given in S1 of the Supplementary file. In these structures, total 45006 Mg2+ ions are found, out of which 18223 are attached to nucleic acids. Out of these 18223 Mg2+ ions which are bound to nucleic acids, some are attached to nucleic acid bases which form base pairs and the rest are attached to bases, including sugar-phosphate groups, which are unpaired. Distributions of Mg-binding sites among these two types of nucleic acid bases, bases which are paired and which are unpaired are shown in Table 1.

Table 1: Distribution of Mg-binding sites among the two types of nucleic acid bases - bases which are paired and which are unpaired

<table>
<tr><th></th><th>Total Numbers</th><th>Sugar</th><th colspan="2">Phosphate</th><th>Base Atoms</th></tr>
<tr><td>Base Pairs</td><td>17951</td><td>479</td><td colspan="2">14527</td><td>2945</td></tr>
<tr><td rowspan="2">Mg Attached with Base Pairs</td><td rowspan="2">13824#</td><td rowspan="2">448</td><td colspan="2">11501</td><td rowspan="2">2537</td></tr>
<tr><td>OP1<br>5867</td><td>OP2<br>6949</td></tr>
<tr><td>Unpaired bases</td><td>8908</td><td>287</td><td colspan="2">7096</td><td>1525</td></tr>
<tr><td>Mg Attached with unpaired Bases</td><td>7474#</td><td>274</td><td colspan="2">6218</td><td>1398</td></tr>
</table>

***#Note: Total Mg involved in sugar + Phosphate + base will not be equal to total Mg count. This is because a single Mg may bind to more than one place and thus will be counted in both the places.***

From Table 1 it is seen that when Mg2+ binds to nucleic acids, it is prone to form base pairs. Figure 3 shows comparative study between Mg-binding sites which are attached to different components of nucleic acids; those are paired and those are unpaired. From Figure 3, it is clear that Mg has a tendency to bind with phosphates than other two components of nucleotide units while it may be in base pair or may not be. In Table 1, the distribution of sites that are associated with two types of phosphate oxygen atoms is also shown in case of base pairs which shows that Mg2+ has a tendency to attach with OP2 atoms more in numbers than with OP1 atoms. It may be noted that OP1 and OP2 are equivalent atoms as the phosphate group is not a chiral center. However, the OP1 is generally named for oxygen atoms facing the minor groove of the double helix while OP2 faces the major groove. Such positioning, however, does not have any meaning where the residue does not belong to a double helix. Close similarity between the occurrence frequencies of Mg ions around the two phosphate oxygen atoms indicate the equivalence.

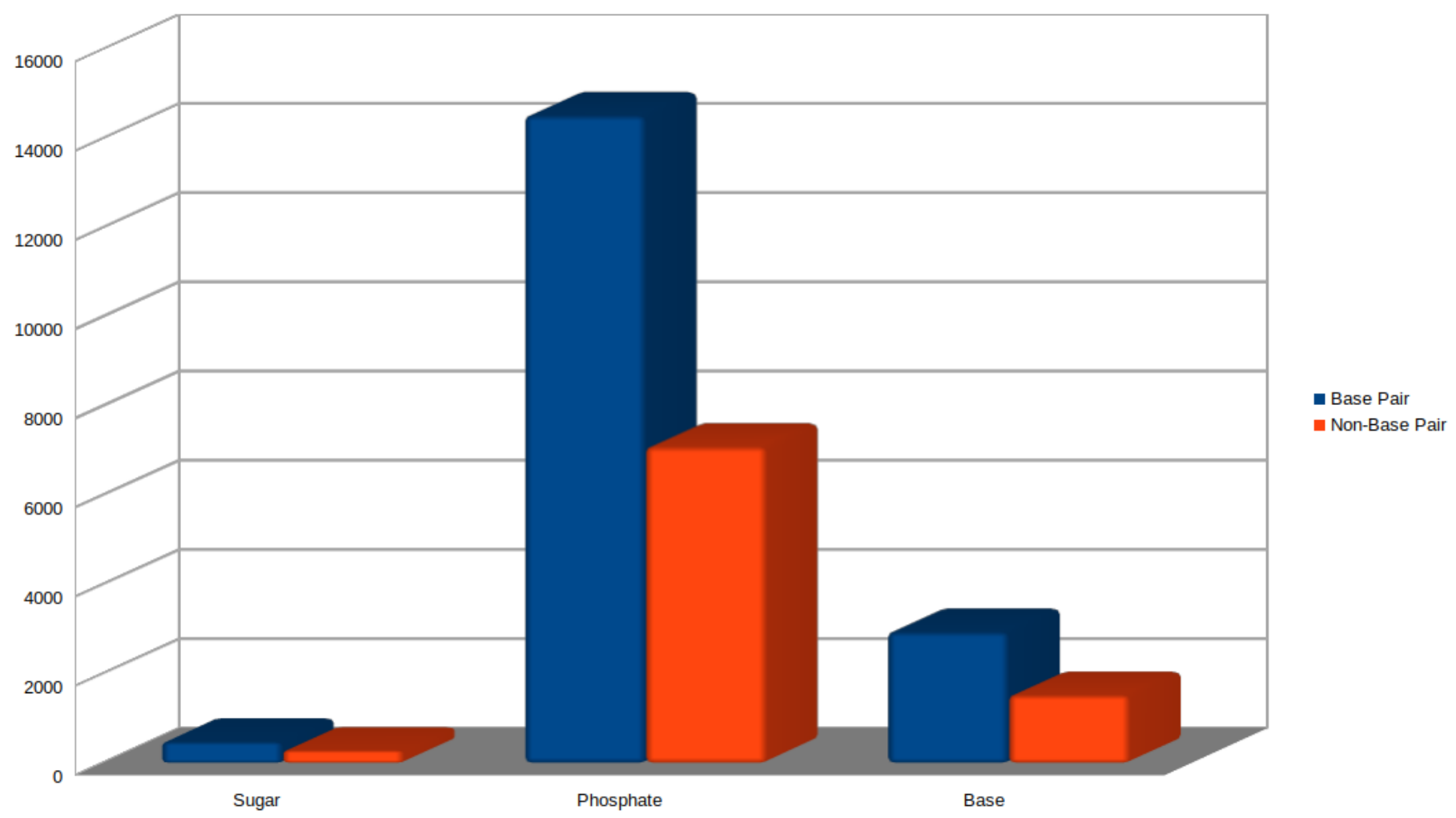

Figure 3: Comparative study between Mg-binding sites which are attached to different components of nucleic acids that are in base pairs and that are not in base pairs

The distribution of Mg-binding sites among the four types of nucleotides in the atoms of the phosphate groups of nucleic acids in paired state is shown in Table 2. In this case, Mg has the highest tendency to bind with adenine (almost 32% of total Mg-binding sites that bind to phosphate) followed by guanine (almost 30% of total Mg-binding sites that bind to phosphate), cytosine (almost 21% of total Mg-binding sites that bind to phosphate) and uracil (almost 15% of total Mg-binding sites that bind to phosphate).

Table 2: Distribution of Mg-binding sites among the four types of nucleotides in the atoms of the phosphate groups of nucleic acids in paired state

| | **A** | **G** | **C** | **U** |
|---|---|---|---|---|
| **Phosphate** | **4678** | 4305 | 3100 | 2212 |

The distributions of Mg-binding sites among the four types of nucleotides in the unpaired and paired base atoms of nucleic acid as well as their atomic distribution are shown in Table 3. In this case, the tendency of Mg to bind with four types of residues in both types, unpaired and paired, is highest in guanine. The distribution of Mg sites in modified atoms in case of unpaired state and their atomic distribution are given in S2 of Supplementary materials. The same for base paired state along with the atomic distribution are shown in S3 of Supplementary materials.

Table 3: Distributions of Mg-binding sites among the four types of nucleobases in the unpaired and base paired state

<table>
<tr><th rowspan="4">Unpaired Base atoms</th><th colspan="5">A</th><th colspan="6">G</th><th colspan="4">C</th><th colspan="4">U</th></tr>
<tr><td colspan="5">146</td><td colspan="6">781</td><td colspan="4">132</td><td colspan="4">428</td></tr>
<tr><th>N1</th><th>N3</th><th>N6</th><th>N7</th><th>N9</th><th>N1</th><th>N2</th><th>N3</th><th>N7</th><th>N9</th><th>O6</th><th>N1</th><th>N3</th><th>N4</th><th>O2</th><th>N1</th><th>N3</th><th>O2</th><th>O4</th></tr>
<tr><td>23</td><td>11</td><td>11</td><td>101</td><td>0</td><td>33</td><td>28</td><td>3</td><td>219</td><td>0</td><td>498</td><td>1</td><td>10</td><td>9</td><td>112</td><td>2</td><td>19</td><td>45</td><td>362</td></tr>
<tr><th rowspan="2">Base paired base atoms</th><td colspan="5">169</td><td colspan="6">1870</td><td colspan="4">123</td><td colspan="4">752</td></tr>
<tr><td>12</td><td>18</td><td>8</td><td>128</td><td>3</td><td>35</td><td>21</td><td>22</td><td>587</td><td>2</td><td>1203</td><td>0</td><td>17</td><td>20</td><td>86</td><td>0</td><td>11</td><td>33</td><td>708</td></tr>
</table>

From Table 3, it is seen that in the case of unpaired state, total Mg-binding sites is 1487 and Mg2+ has the highest tendency to attach to O6 atom (almost 33% of total Mg-binding sites that bind to nucleobases) of guanines followed by its N7 atom (almost 14% of total Mg-binding sites that bind to nucleobases). Mg2+ also has the highest tendency to bind with N7 (almost 7% of total Mg-binding sites that bind to nucleobases), O4 (almost 24% of total Mg-binding sites that bind to nucleobases) and O2 (almost 7% of total Mg-binding sites that bind to nucleobases) atoms in case of adenine, uracil and cytosine respectively. The O2 atom of Uracil is found to be not favorable for Mg2+ ion for binding. Similarly the N3 atoms of Adenine and Guanine do not bind to Mg2+ with high preference. This observation is possibly the crowding effect by sugar moiety. The primary and secondary amino groups of the four bases are also found to be interacting with Magnesium ions, although with quite low frequency. In unpaired state, Mg2+ does not bind with the N9 atom of adenine and guanine as this nitrogen is fully saturated with three bonds. Surprisingly the N3 atom of Cytosine shows preference towards Mg binding similar to N3 of Uracil although N3 of Uracil is secondary amino nitrogen with a hydrogen attached to it while N3 of Cytosine is imino nitrogen.

From Table 3, it is also seen that in paired state, total Mg-binding sites is 2914 and Mg2+ has the highest tendency to attach with the O6 atom of guanines (almost 41% of total Mg-binding sites that bind to nucleobases) followed by the N7 atom (almost 20% of total Mg-binding sites that bind to nucleobases). Mg2+ also has the highest tendency to bind with N7 (almost 4% of total Mg-binding sites that bind to nucleobases), O4 (almost 24% of total Mg-binding sites that bind to nucleobases) and O2 (almost 3% of total Mg-binding sites that bind to nucleobases) atoms in case of adenine, uracil and cytosine respectively in paired state. [In both unpaired and paired states, all the percentage values are calculated by using data in Table 1 and Tables 3]

Table 3 also shows that although Mg2+ prefers to bind with O6 atom of guanine followed by its N7 atom in both unpaired and paired state, comparatively the preference is more in paired state.

The distribution of Mg-binding sites among the four types of nucleotide residues in the atoms of the sugar edges of nucleic acids in unpaired and paired states is also shown in Table 4. In both unpaired and paired state the tendency of Mg2+ to bind with four types of residues is highest in adenine (almost 29% and 32% of total Mg-binding sites that bind to sugar edge respectively) followed by guanine (almost 22.9% and 30% of total Mg-binding sites that bind to sugar edge respectively), cytosine (almost 22.6% and 24% of total Mg-binding sites that bind to sugar edge respectively) and uracil (almost 20% and 13% of total Mg-binding sites that bind to sugar edge respectively). The distribution of Mg2+ sites in modified residues in the sugar edges of nucleic acids in unpaired state is given in **S4** Supplementary materials. [In both unpaired and paired states, all the percentage values are calculated by using data in Tables 1 and 4.]

The atomic distribution of Mg-binding sites among the four types of residues in the sugar atoms in unpaired and paired states is also shown in Table 4. This table shows that in both cases Mg2+ has the highest tendency to bind with O2* atom of sugar edges followed by the O3* atoms in all the four types of cases. S4 of Supplementary materials shows the distribution of Mg2+ sites in modified residues in the atoms of the sugar edges of nucleic acids that do not form base pairs. This probably explains high importance of base pairs stabilized by hydrogen bonds involving O2* as either hydrogen bond acceptor as well as donor. In this state, Mg does not interact with the O4* atom of uracils whereas it interacts with this atom in other three nucleobases in sugar edges.

Table 4: Distribution of Mg-binding sites among the four types of nucleobases in the sugar edges of nucleic acids along with the atomic distribution in unpaired and paired states

| **Sugar atoms of unpaired residues** | **A** | | | | **G** | | | | **C** | | | | **U** | | | |
|---|---|---|---|---|---|---|---|---|---|---|---|---|---|---|---|---|
| | **83** | | | | 66 | | | | 65 | | | | 57 | | | |
| | **O2*** | **O3*** | **O4*** | **O5*** | **O2*** | **O3*** | **O4*** | **O5*** | **O2*** | **O3*** | **O4*** | **O5*** | **O2*** | **O3*** | **O4*** | **O5*** |
| | **47** | **23** | **3** | **10** | **46** | 16 | 1 | 3 | **45** | 11 | 6 | 3 | **43** | 10 | 1 | 3 |
| **Sugar atoms of base paired residues** | **151** | | | | **145** | | | | **117** | | | | **63** | | | |
| | **94** | 39 | 5 | 13 | **78** | 50 | 2 | 15 | **68** | 39 | 5 | 5 | **38** | 15 | 0 | 10 |

A comparative study between Mg-binding sites which are attached to O2* and O3* atoms of sugar edges of nucleic acids that are not in base pairs are depicted in Figure 4. This figure shows the tendency of Mg to bind with O2* atoms of the sugar edges of adenine and uracil are the highest and lowest respectively, which are almost 16% and 15% of the Mg-binding sites that bind to atoms of the sugar edges. On the other hand, the tendency of Mg2+ to bind with O3* of four types of nucleotides are very poor. The phosphodiester oxygen atoms do not bind to Mg possibly due to close proximity of more electronegative OP1 and OP2 atoms. It is highest in adenine (almost 8% of the Mg-binding sites that bind to atoms of the sugar edges) and lowest in uracil (almost 3% of the Mg-binding sites that bind to atoms of the sugar edges).

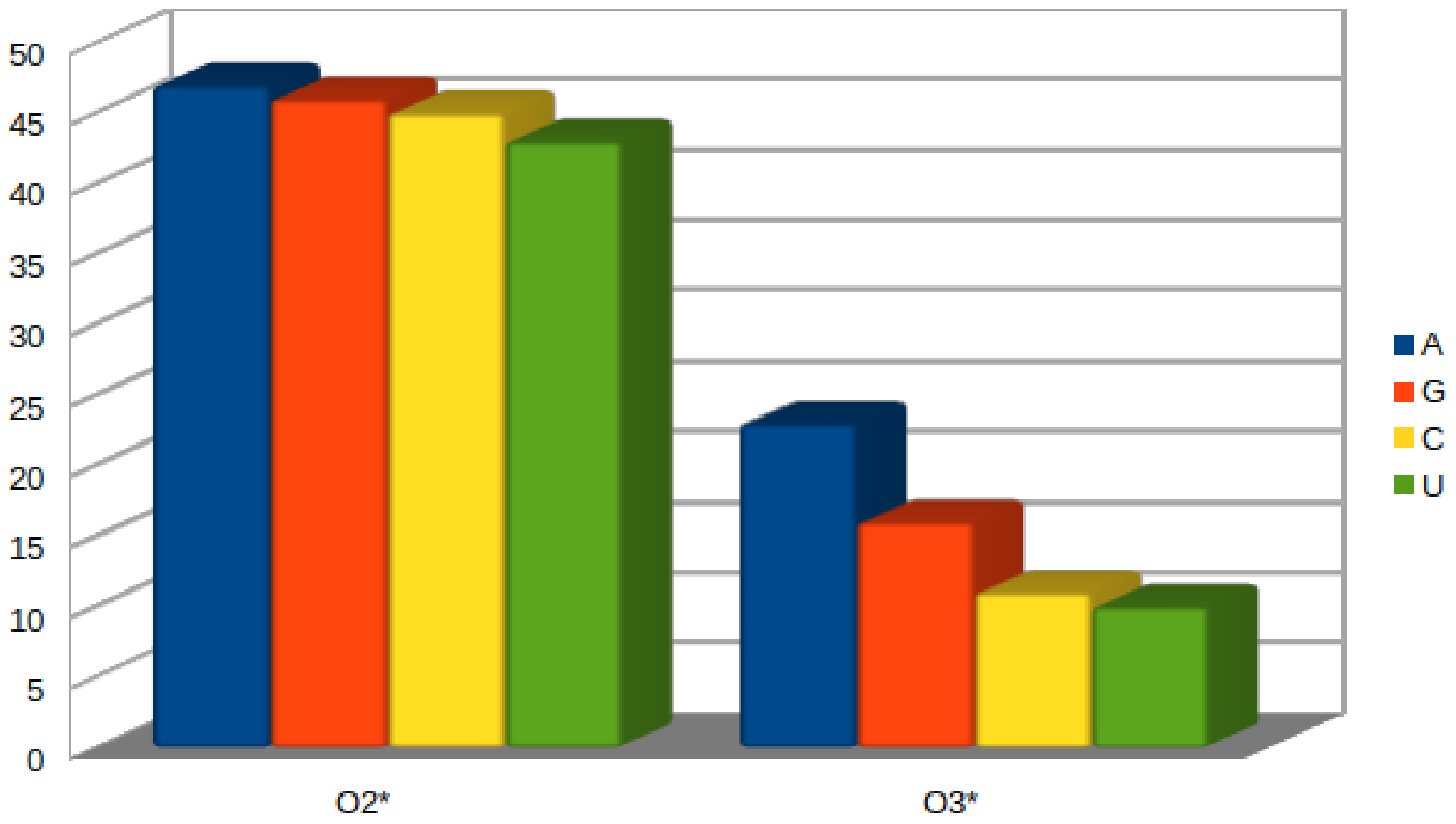

Figure 4: Comparative study between Mg-binding sites which are attached to O2* & O3* atoms of sugar edges of nucleic acids that are not in base pairs

A comparative study between Mg-binding sites which are attached to O2* and O3* atoms of sugar of nucleic acids that are in base paired are depicted in S5 of the Supplementary materials (Figure 1). This figure shows the tendency of Mg2+ to bind to O2* atoms of the sugar of adenine and uracil are the highest and lowest respectively, which are almost 20% & 8% of the Mg-binding sites that bind to atoms of the sugar edges. On the other hand, the tendency of Mg to bind with O3* atom of guanine is the highest in comparison to the other O3* binding nucleobases which is almost 10% of the Mg-binding sites that bind to atoms of the sugar edges. Again, uracil is the lowest in binding Mg to its O3* atoms, which is almost 3% of the Mg-binding sites that bind to atoms of the sugar edges.

The base pairs which are found to form O2*...Mg interaction are A:C-s:sC, A:G-s:sT, A:G-s:wC, G:C-S:WT, C:G-s:hT, G:U-s:hC, G:U-S:WC, G:A-S:HT, G:A-S:WT, C:C-s:hT, C:U-S:WT, A:A-s:hT, A:A-S:HC, A:U-s:wT, G:G-S:SC. All these base pairs are utilizing their sugar edges i.e., hydrogen bond involving sugar O2* is mandatory for these base pair formations.

### III.2 Interaction of Mg2+ ions with Watson-Crick base pairs & non-Watson-Crick base pairs

The distribution of Mg-binding sites among the four types of nucleotides in the Watson-Crick (canonical) and non Watson-Crick (non canonical) base pairs of nucleic acids is shown in Table 5. Among the four natural nucleotides, only guanine is seen to interact with Mg2+ more frequently when it forms a canonical base pair and only adenine is seen to interact more frequently with Mg2+ when it forms a non-canonical base pair. The number distribution further indicates equal importance of canonical and non-canonical base pairing in RNA structure.

Table 5: Distribution of Mg-binding sites among the four types of nucleobases when the bases are in the Watson-Crick and non Watson-Crick base pairs.

| | **A** | **G** | **C** | **U** |
|---|---|---|---|---|
| **Watson-Crick** | 1119 | 3253 | 2826 | 948 |
| **Non Watson-Crick** | 3535 | 2881 | 347 | 1964 |

Next, we shall show that not only canonical base pairs, non-canonical base pairs also have contribution in Mg-binding. For this, we have computed the number of Mg-binding sites that are associated with each of the following canonical and non-canonical base pairs along with the bonding residues and binding atoms ( Table 6). Here, the total number of Mg-bound canonical and non-canonical base pairs are found to be 971 and 1769 respectively. From this data it is clear that the non-canonical base pairs are more dependent on Mg binding. From Table 6, it is seen that comparatively adenine and cytosine are not so good for Mg-binding and guanine is best for the same followed by uracil in case of canonical base pairing.

Table 6 also shows the Mg-binding statistics related to 21 types of non-canonical base pairs (where the count of Mg binding sites is more than or equal to 10). From these statistics, it appears that for binding of Mg2+ ions, cytosine is not so good for binding of Mg2+ ions and again guanine supersedes the other nucleobases. All figures (created in PyMol) related to these base pairs which are shown in Table 6 are presented in S6 of the Supplementary materials.

Table 6 again shows that the N7 and O6 atoms of guanine are involved in interaction with Mg2+ ions along with its N1, N2, N3, O2*, O3*, O4* and O5* atoms in both of canonical and non-canonical base pairing like in G:C-W:WC, G:U-W:WC, G:A-S:HT, G:G-S:ST, G:G-W:HC, G:U s:hC, G:U-S:WC, G:C-W:ST, G:C-S:WT, G:A W:WC, G:A S:WT and G:G-W:HT base pairs. All these atoms interact with Mg2+ ions in G:C-W:WC and G:U-W:WC base pairs. Some of these atoms interact with Mg2+ ions in other non-canonical base pairing of guanine.

From Table 6 it is also seen that in all the base pairs, there is a tendency of Mg2+ ions to bind with sugar or phosphate atoms than the base atoms except some of the non-canonical base pairs like U:A-W:HT, G:G W:HC, C:A W:HT, G:U S:WC, G:C W:ST, G:A s:wT and G:G W:HT.

Table 6: Statistics of Mg-binding sites related to canonical & non-canonical base pairs

| **Base Pair Name** | **No. of Base Pairs** | **Bonding Residue** | **Binding Atoms** | **No. of Mg-binding Sites** |
|---|---|---|---|---|
| G:C W:WC | 101047 | G | Base Atoms: N1/ N2/ N3/ N7/ O6 | 655 |
| | | | Sugar-Phosphate Atoms: O2*/ O3*/ O4*/ O5*/ OP1/ OP2/ OP3 | 2598 |
| C:G W:WC | | C | Base Atoms: N3/ N4/ O2 | 55 |
| | | | Sugar-Phosphate Atoms: O2*/ O3*/ O4*/ O5*/ OP1/ OP2 | 2771 |
| A:U W:WC | 39915 | A | Base Atoms: N1/ N3/ N6/ N7/ N9 | 37 |
| | | | Sugar-Phosphate Atoms: O2*/ O3*/ O4*/ O5*/ OP1/ OP2/ OP3 | 1082 |
| U:A W:WC | | U | Base Atoms: N3/ O2/ O4 | 213 |
| | | | Sugar-Phosphate Atoms: O2*/ O3*/ O4*/ O5*/ OP1/ OP2 | 735 |
| G:U W:WC | 17275 | G | Base Atoms: N1/ N2/ N3/ N7/ O6 | 320 |
| | | | Sugar-Phosphate Atoms: O2*/ O3*/ O4*/ O5*/ OP1/ OP2 | 376 |
| U:G W:WC | | U | Base Atoms: N3/ O2/ O4 | 277 |
| | | | Sugar-Phosphate Atoms: O2*/ O3*/ O4*/ O5*/ OP1/ OP2 | 468 |
| G:A S:HT# | 10156 | G | Base Atoms: N1/ N2/ N3/ N7/ N9/ O6 | 501 |
| | | | Sugar-Phosphate Atoms: O2*/ O3*/ O5*/ OP1/ OP2 | 599 |
| U:A W:HT# | 4757 | U | Base Atoms: O2/ O4 | 169 |
| | | | Sugar-Phosphate Atoms: O2*/ O5*/ OP1/ OP2 | 161 |
| G:G S:ST# | 547 | G | Base Atoms: N2/ O6 | 83 |
| | | | Sugar-Phosphate Atoms: OP1/ OP2 | 186 |
| G:G W:HC# | 178 | G | Base Atoms: N1/ N2/ N7/ O6 | 71 |
| | | | Sugar-Phosphate Atoms: O3*/ O5*/ OP1/ OP2 | 16 |
| G:U s:hC# | 1036 | G | Base Atoms: N1/ O6 | 54 |
| | | | Sugar-Phosphate Atoms: O2*/ O5*/ OP1/ OP2 | 101 |
| C:A W:HT# | 1291 | C | Base Atoms: O2 | 44 |
| | | | Sugar-Phosphate Atoms: O2*/ O3*/ O4*/ OP1/ OP2 | 36 |
| A:G S:SC# | 330 | A | Base Atoms: N7 | 36 |
| | | | Sugar-Phosphate Atoms: OP1/ OP2 | 86 |
| A:A s:hT# | 562 | A | Base Atoms: N7 | 35 |
| | | | Sugar-Phosphate Atoms: O2*/ OP1/ OP2 | 92 |
| G:U S:WC# | 196 | G | Base Atoms: N7/ O6 | 30 |
| | | | Sugar-Phosphate Atoms: O2*/ OP2 | 3 |
| U:U W:WC# | 2553 | U | Base Atoms: O4 | 29 |
| | | | Sugar-Phosphate Atoms: O2*/ OP1/ OP2 | 251 |
| G:C W:ST# | 150 | G | Base Atoms: N7 | 23 |
| | | | Sugar-Phosphate Atoms: O4*/ OP1/ OP2 | 15 |
| U:G S:WC# | 200 | U | Base Atoms: O4 | 22 |
| | | | Sugar-Phosphate Atoms: OP1 | 84 |
| G:C S:WT# | 145 | G | Base Atoms: N1/ N7 | 18 |
| | | | Sugar-Phosphate Atoms: O2*/ OP1/ OP2 | 29 |
| G:A s:wT# | 962 | G | Base Atoms: N7/ O6 | 17 |
| | | | Sugar-Phosphate Atoms: OP1/ OP2 | 10 |
| A:C s:sC# | 932 | A | Base Atoms: N7 | 16 |
| | | | Sugar-Phosphate Atoms: O2*/ O3*/ OP1/ OP2 | 127 |
| U:C W:WC | 579 | U | Base Atoms: O2 | 12 |
| | | | Sugar-Phosphate Atoms: OP1/ OP2 | 48 |
| | | C | Base Atoms: O2 | 12 |
| | | | Sugar-Phosphate Atoms: OP1/ OP2 | 25 |
| G:A W:WC# | 2688 | G | Base Atoms: N3/ N7/ O6 | 12 |
| | | | Sugar-Phosphate Atoms: O5*/ OP1/ OP2 | 193 |
| G:A S:WT# | 572 | G | Base Atoms: N7/ O6 | 12 |
| | | | Sugar-Phosphate Atoms: OP1/ OP2 | 42 |
| G:G W:HT# | 51 | G | Base Atoms: N1/ N2/ N7/ O6 | 10 |
| | | | Sugar-Phosphate Atoms: OP2 | 4 |
| A:A H:HT# | 1627 | A | Base Atoms: N1/ N3/ N6 | 10 |
| | | | Sugar-Phosphate Atoms: O2*/ OP1/ OP2 | 196 |

***#Note: In these base pairs, their duals are not shown because in these dual base pairs like A:G-H:ST, A:U-H:WT etc., the numbers of Mg-binding sites are less than 10. Only the base pairs where the numbers of Mg-binding sites are greater than or equal to 10 are shown in this Table.***

Let us now turn our attention to edges, i.e. which edge is preferred for *base-pairs* when Mg is forming a coordination. Details of such binding edges for adenine involved base pairs along with the Mg2+ binding atom name is presented in Table 7 and the same for cytosine and uracil is presented in Table 8. Table 9 shows such specifications in guanine. In these tables, both canonical and non-canonical base pairs are considered. Tables 7-9 show that the tendency of Mg2+ ions to bind with N1 atoms of adenine when adenine uses its Hoogsteen edge for base pairing is maximum (almost 26%) and Mg2+ ions bind only with O6 (almost 48% of total binding in the Hoogsteen edge) atoms of guanine in the Hoogsteen edge. These three tables also show that Mg2+ ions prefer to bind with atoms of guanine in Sugar edges (almost 85% of total binding in the Sugar edge) followed by atoms of adenine (almost 11% of total binding in the Sugar edge). The tendency to bind with the atoms of cytosine and uracil in the sugar edge is not so significant. Mg2+ ions prefer to bind with atoms of guanine in Watson-Crick edges (almost 56% of total binding in the Watson-Crick edge) followed by atoms of uracil (almost 35% of total binding in the Watson-Crick edge).

Table 7: Statistics related to details of binding atoms in adenine

| **When Hoogsteen edge forms base pair** | | | **When Sugar edge forms base pair** | | **When Watson-Crick edge forms base pair** | | | | |
|---|---|---|---|---|---|---|---|---|---|
| **N1** | **N3** | **N6** | **N1** | **N7** | **N1** | **N3** | **N6** | **N7** | **N9** |
| 7 | 6 | 1 | 1 | 93 | 4 | 12 | 7 | 35 | 3 |

Table 8: Statistics related to details of binding atoms in Cytosine & Uracil.

| Cytosine | | | | Uracil | | | |
|---|---|---|---|---|---|---|---|
| **When Sugar edge forms base pair** | **When Watson-Crick edge forms base pair** | | | **When Sugar edge forms base pair** | **When Watson-Crick edge forms base pair** | | |
| **N3** | **N3** | **N4** | **O2** | **O4** | **N3** | **O2** | **O4** |
| 3 | 14 | 20 | 86 | 36 | 11 | 33 | 663 |

Table 9: Statistics related to details of binding atoms in guanine

| When Hoogsteen edge forms base pair | When Sugar edge forms base pair | | | | | | When Watson-Crick edge forms base pair | | | | |
|---|---|---|---|---|---|---|---|---|---|---|---|
| **O6** | **N1** | **N2** | **N3** | **N7** | **N9** | **O6** | **N1** | **N2** | **N3** | **N7** | **O6** |
| 13 | 10 | 4 | 1 | 257 | 2 | 468 | 25 | 17 | 21 | 328 | 720 |

From Table 7 it is seen that although the N1 atom of adenine is involved in formation of hydrogen bonds in its Watson-Crick edge, still Mg2+ ions are in close proximity to the N1 atom which is depicted in Figure 5. In Figure 5(a), the base pair A:U-W:WC [Chain ID: 1, Residue Names: A, U, Residue Nos.: 1014, 1148; Mg: Chain ID: 1, Residue Name: MG, Residue No.: 3120] of the structure 7OT5.pdb is shown where the N1 atom of the adenine is taking part in interaction with Mg2+ ion as well as formation of hydrogen bonds. Figure 5(b) shows the same for trans orientation of A:U-W:W base pair [Chain ID: 2, Residue Names: A, U, Residue Nos.: 397, 37; Mg: Chain ID: 2, Residue Name: MG, Residue No.: 1680] of the structure 7OIF.pdb.

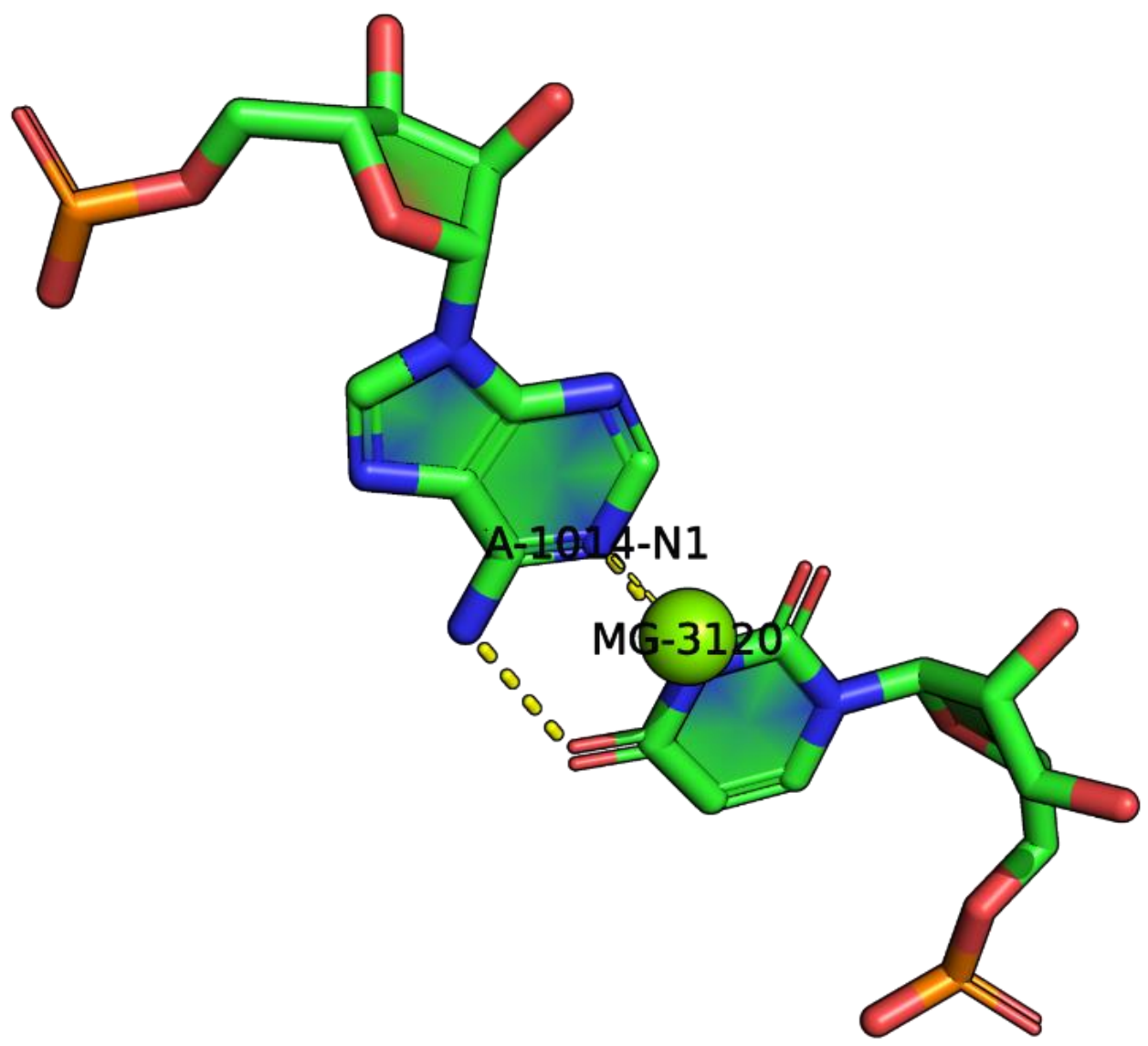

(a) PDB ID: 7OT5 - Base Pair: A:U-W:WC [Chain ID: 1, Residue Names: A, U, Residue Nos.: 1014, 1148]; Mg [Chain ID: 1, Residue Name: MG, Residue No.: 3120]

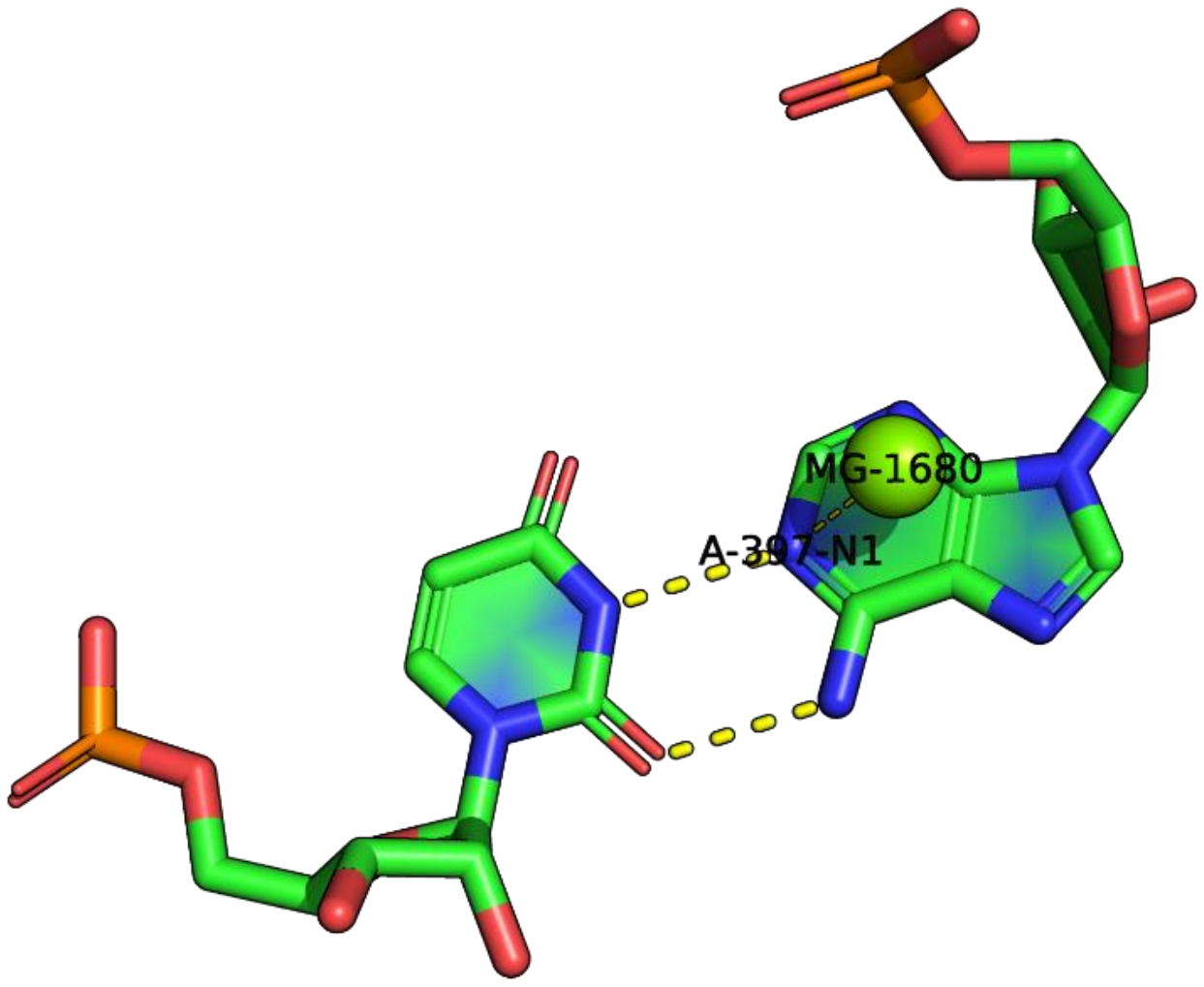

(b) PDB ID: 7OIF - Base Pair: A:U-W:WT [Chain ID: 2, Residue Names: A, U, Residue Nos.: 397, 37]; Mg [Chain ID: 2, Residue Name: MG, Residue No.: 1680]

Figure 5: Binding of Mg2+ with the atom N1 of Adenine when its Watson-Crick edge is involved in base pairing

Table 8 also shows the same for the pyrimidine bases. In this table, it is seen that although N3 atoms of cytosine and uracil are also involved in formation of hydrogen bonds their Watson-Crick edges, still Mg2+ ions are in close proximity to the N3 atoms which are depicted in Figures 6 and 7 respectively. In Figure 6, the base pair C:G-W:WC [Chain ID: 1, Residue Names: C, G, Residue Nos.: 1349, 1382; Mg: Chain ID: 1, Residue Name: MG, Residue No.: 3265] of the structure 7OIF.pdb is shown where the N3 atom of the cytosine is taking part in interaction with Mg2+ ion as well as formation of hydrogen bonds.

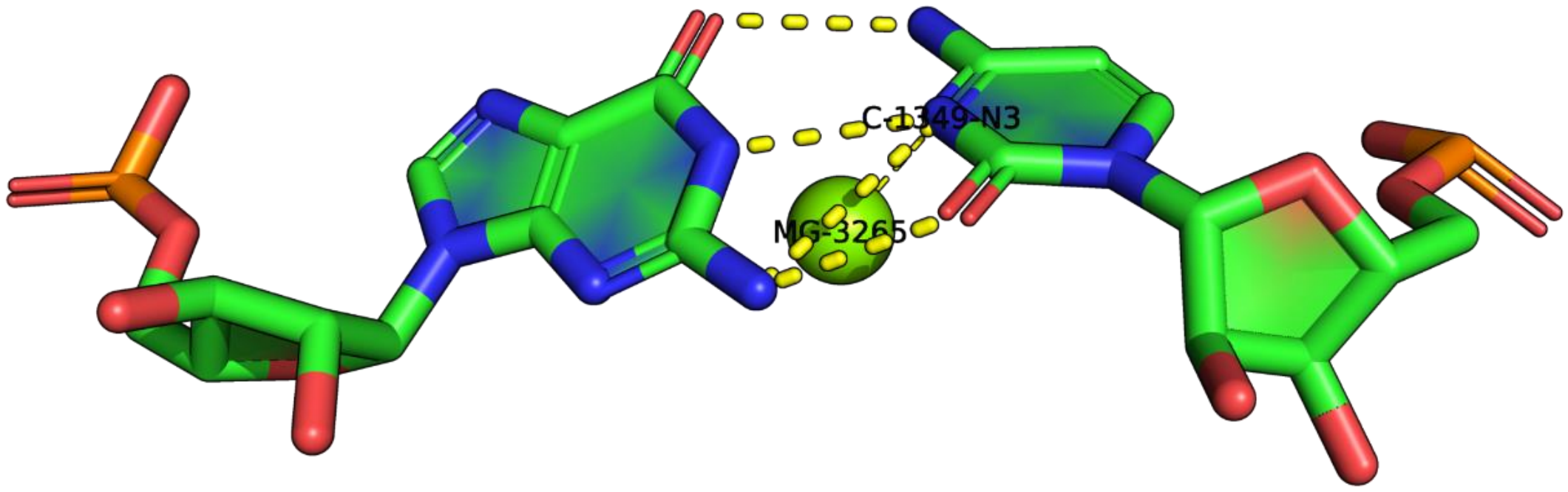

Figure 6: Binding of Mg2+ with the atom N3 of Cytosine when its Watson-Crick edge is involved in base pairing - PDB ID: 7OIF - Base Pair: C:G-W:WC [Chain ID: 1, Residue Names: C, G, Residue Nos.: 1349, 1382]; Mg [Chain ID: 1, Residue Name: MG, Residue No.: 3265]

In Figure 7, the base pair U:G-W:WC [Chain ID: 1, Residue Names: U, G, Residue Nos.: 1851, 1891; Mg: Chain ID: 1, Residue Name: MG, Residue No.: 3101] of the structure 7OT5.pdb is shown where the N3 atom of the uracil is involved in formation of hydrogen bonds and interacts with Mg2+ ion.

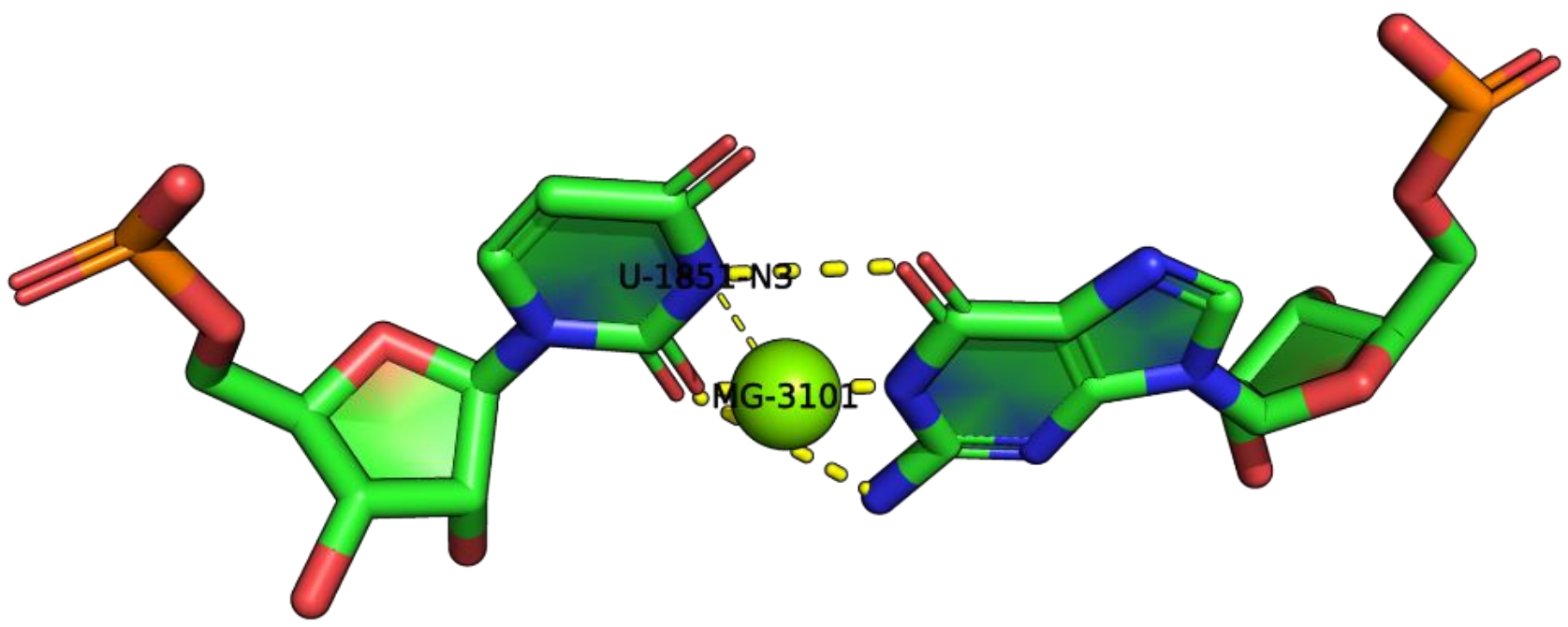

Figure 7: Binding of Mg2+ with the atom N3 of Uracil when its Watson-Crick edge is involved in base pairing - PDB ID: 7OT5 - Base Pair: U:G-W:WC [Chain ID: 1, Residue Names: U, G, Residue Nos.: 1851, 1891]; Mg [Chain ID: 1, Residue Name: MG, Residue No.: 3101]

Another base pair, G:G-W:HC [Chain ID: A, Residue Names: G, G, Residue Nos.: 299, 566; Mg: Chain ID: A, Residue Name: MG, Residue No.: 1576] of the structure 1N32.pdb is shown where the Mg2+ ion is interacting with the O6 atoms of both participating bases. Among these two O6 atoms, one O6 atom is involved in formation of hydrogen bonds. This case is depicted in Figure 8. It may be noted that this base pairing is an integral part of telomere structure at the 3'-overhang region of chromosomal DNA. This base pair has been observed quite frequently in the available RNA structures as well. There were several studies using quantum chemical calculations to understand strength and stability of this base pair by performing geometry optimization [21-22]. Such geometry optimizations of the isolated base pairs, however, lead to another type of base pairing, namely W:WT. This is presumably due to the close approach of two negatively charged O6 atoms in such base pairing. In DNA telomere there are four Guanine bases forming quadruplex and is often stabilized by K+ or Na+ at the center. In the RNA structures also we noted the majority of G:G W:HC base pairs are being additionally stabilized by Mg2+ ions.

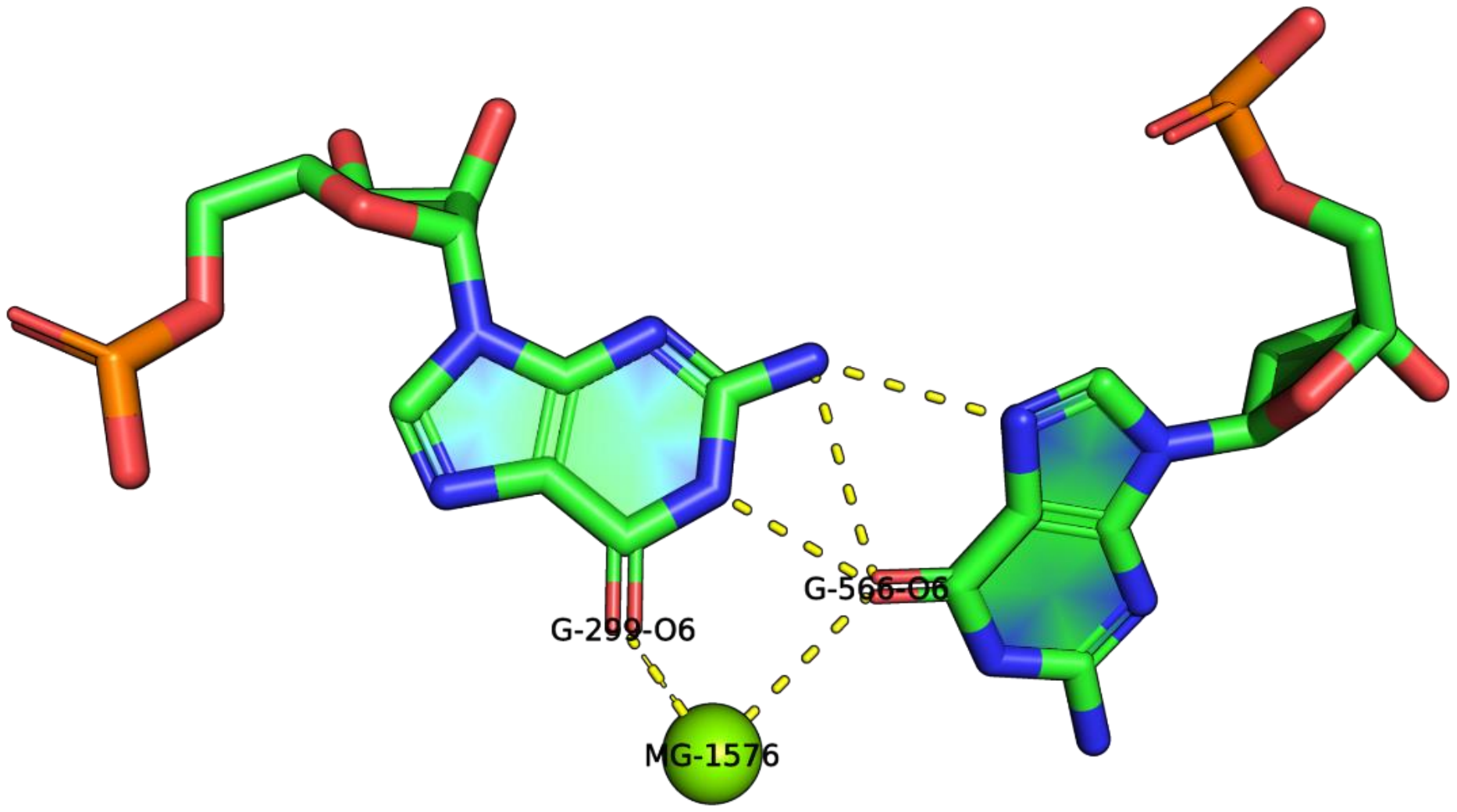

Figure 8: Binding of Mg2+ with the atom O6 of Guanine when its Hoogsteen edge is involved in base pairing – PDB ID: 1N32 - Base Pair: G:G-W:HC [Chain ID: A, Residue Names: G, G, Residue Nos.: 299, 566]; Mg [Chain ID: A, Residue Name: MG, Residue No.: 1576]

A comparative study of edge preferences for base pairs when there is influence of Mg are depicted in Figure 9. It shows that guanine prefers to offer its sugar edge for base pairing when Mg2+ ions are in proximity than in Watson-Crick and Hoogsteen edges whereas in case of adenine, the preference of such base pairing is for Hoogsteen edges.

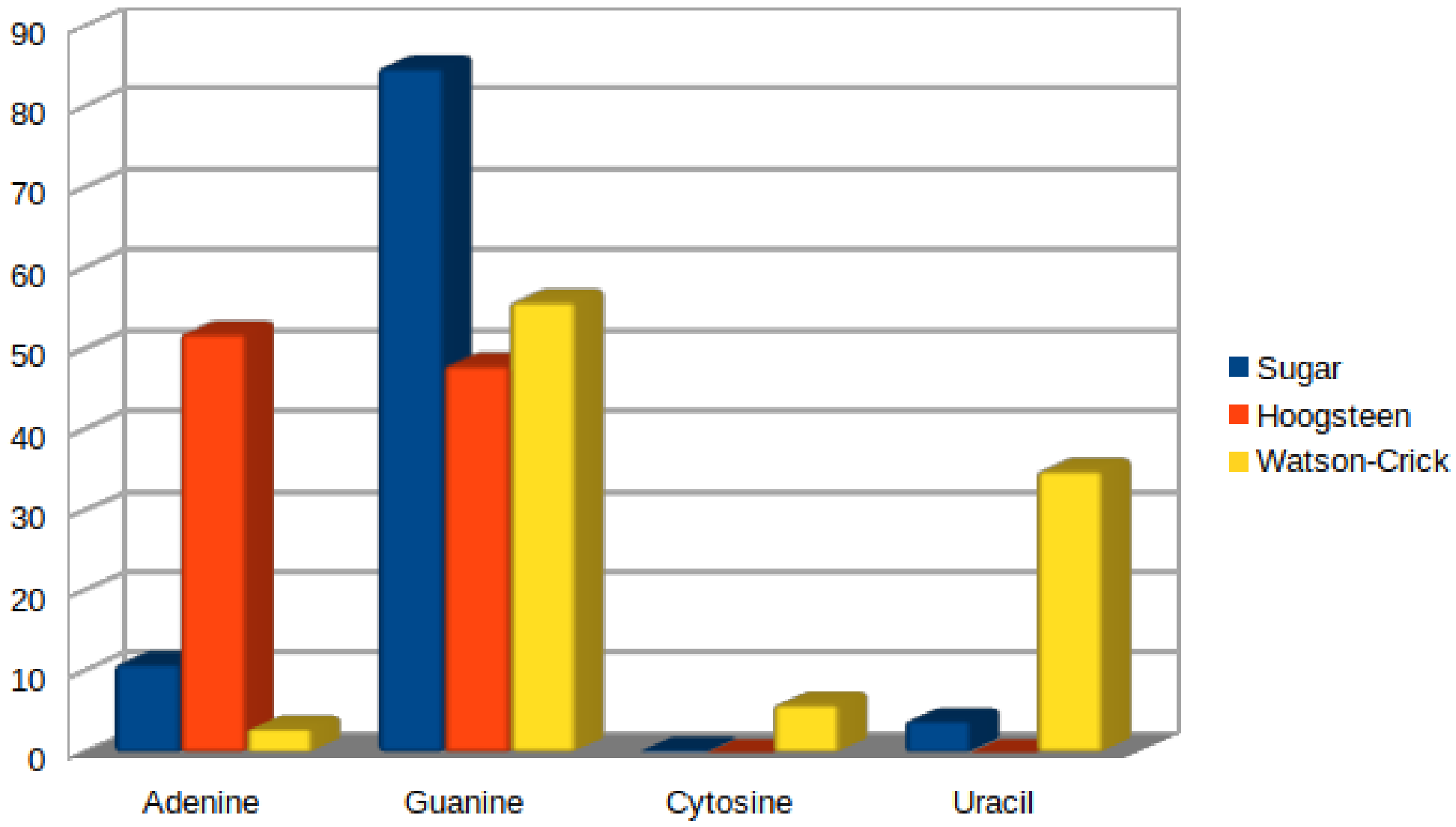

Figure 9: Contribution of edges to the base pairs for Mg-binding in nucleic acids.

The distances between the binding atoms and the Mg2+ ions are also computed. The mean, median, first quartile value, second quartile value and the standard deviation of such distances are presented in Table 10. From Table 10 it is seen that here, the mean is equal to median and the standard deviation shows that such distances do not deviate too much from the mean. The values of q3-median & median-q1 are 0.022 & 0.058 respectively which implies that the distribution is almost symmetric. As we have taken a very short distance cutoff, we are getting the statistics which show very close range interactions.

Table 10: Distance of Mg to binding atom - mean, median, q1, q3, standard deviation

| **Mean** | **Median** | **1st Quartile (q1)** | **3rd Quartile (q3)** | **Standard Deviation** |
|---|---|---|---|---|
| 2.179 | 2.179 | 2.121 | 2.201 | 0.161 |

Stabilities of the Mg2+-bound base pairs, in terms of the E-values are computed in order to have an idea about how stable these base pairs are and whether binding of Mg2+ ions makes any change to the stability of these base pairs. For this, we have calculated the average stability in presence of Mg2+ ion and the average stability in absence of Mg2+ ion for the base pairs. Here, the stability of base pairs are considered to be very strong if the E-values are closer to zeroes. Different statistics related to base pairs like mean and standard deviation of E-values and number of such base pairs for each type of base pairs are presented in Table 11. It shows that comparatively G:C-W:WC base pair is the most frequent and stable base pair when Mg2+ ion is not attached to this base pair. Now, Mg2+ ions may attach to the guanine or cytosine or both in cases of residue type base or sugar. Whatever may be the case, interaction of Mg2+ ions with this base pair slightly reduces the mean E-value that means the strength of this base pair increases. The standard deviation of E-values, in case of base pairs, shows more deviation from mean due to presence of Mg2+ ions with this base pair than when Mg2+ ions are not attached to this base pair. This is presumably due to reduction of dipole moment of guanine upon Mg2+ binding as one of the main component of interaction between guanine and cytosine in W:WC type of base pairing is dipole-dipole interaction with large dipole moments of both guanine and cytosine.

Table 11: Different statistics related to E-values of base pairs. (*For measurement of stability. The lower the E-value more stable it is. Follow Equation 1 for its calculation.*)

| **Base Pair** | **Binding Location** | **Binding Residues** | **Mean E-value** | | **@Stdev of E-values** | | **Count** | |
|---|---|---|---|---|---|---|---|---|
| | | | **When Mg** | **When Mg** | **When Mg is** | **When Mg** | **When Mg is** | **When Mg** |

| | | | is in proximity | is not in proximity | not in proximity | is in proximity | not in proximity | is in proximity |
|---|---|---|---|---|---|---|---|---|
| G:C-W:WC | Base | C | 0.27 | 0.31 | 0.19 | 0.23 | 101047 | 55 |
| | | G | 0.29 | | 0.19 | 0.16 | | 655 |
| | Sugar | C | 0.31 | 0.29 | 0.19 | 0.10 | | 81 |
| | | G | 0.31 | 0.30 | 0.19 | 0.19 | | 80 |
| | | | | | | | | |
| G:U-W:WC | Base | G | 0.26 | 0.21 | 0.24 | 0.17 | 17275 | 320 |
| | | U | 0.26 | 0.22 | 0.24 | 0.15 | | 672 |
| | | | | | | | | |
| G:U-s:hC | Base | G | 0.89 | 1.04 | 0.24 | 0.25 | 1036 | 54 |
| | | | | | | | | |
| G:U-S:WC | Base | G | 0.65 | 0.66 | 0.25 | 0.20 | 196 | 30 |
| | | | | | | | | |
| A:U-W:WC | Base | A | 0.32 | 0.35 | 0.22 | 0.24 | 39915 | 37 |
| | | U | 0.32 | 0.37 | 0.22 | 0.20 | | 213 |
| | Sugar | A | 0.32 | 0.28 | 0.22 | 0.11 | | 30 |
| | | U | 0.32 | 0.37 | 0.22 | 0.28 | | 19 |
| | | | | | | | | |
| G:G-S:ST | Base | G | 0.73 | 0.68 | 0.31 | 0.25 | 547 | 83 |
| | | | | | | | | |
| G:G-W:HC | Base | G | 0.51 | 0.34 | 0.36 | 0.18 | 178 | 71 |
| | | | | | | | | |
| U:U-W:WC | Base | U | 0.41 | 0.35 | 0.27 | 0.23 | 2553 | 29 |
| | | | | | | | | |
| A:A-s:hT | Base | A | 0.48 | 0.49 | 0.27 | 0.32 | 562 | 35 |

A statistical report regarding the position of a base pair in the secondary structure of RNA like helix, coil, loop etc. with which Mg2+ has been interacted is also generated. Table 12 shows the same. From Table 12 it is seen that a maximum of base pairs reside within the loop portion of the secondary structure of RNA.

Table 12: Report regarding the position of a base pair in the secondary structure of RNA with which Mg2+ has been interacted

| **No. of Sites in** | | | | |
|---|---|---|---|---|
| **Bulge** | **Helix** | **Loop** | **Coil** | **Isolated Watson-Crick base pair** |
| 92 | 8710 | 12949 | 119 | 2837 |

In the present study, the statistics related to formation of chelates by the Mg2+ ion in nucleic acid structures are also presented. The chelate that is formed by a Mg2+ ion in the structure of nucleic acid (PDB ID: 1VQN) is depicted in Figure 10 which is also created in PyMol.[16]

Residue No.: 1258, Chain ID: 0

Residue No.: 1098, Chain ID: 0

Residue Name: MG, Residue No.: 8094, Chain ID: 0

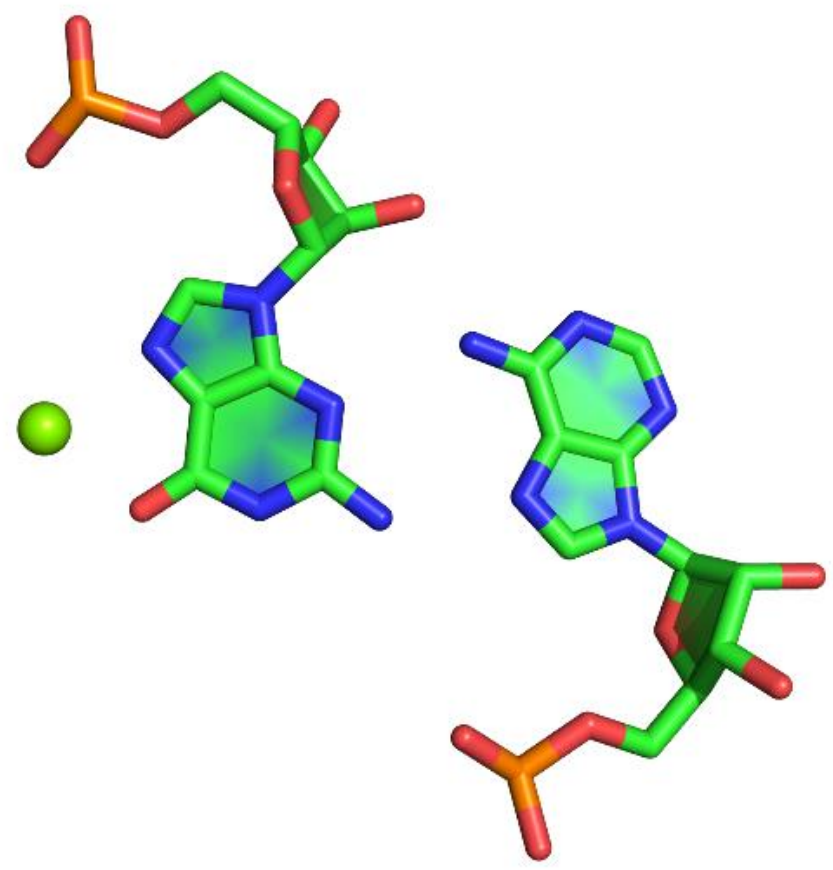

Figure 10: Formation of chelate by a Mg2+ ion - PDB ID: 1VQN [Chain ID: 0, Res Name: MG, Res No.: 8094; Chain ID: 0, Res No.: 1258; Chain ID: 0, Res No. 1098]

We have found 389 chelates in total. The distribution of such sites is shown in Table 13. We found three types of chelates, namely, alpha, beta and gamma where alpha, beta and gamma are as follows:

(i) When the residue type is phosphate or the residue type, the first & the second binding locations all are nucleobases then the atom name is alpha.

(ii) When the residue type is nucleobase and the first & the second binding locations are any one of nucleobase, phosphate & sugar (not both same), the atom is named as beta.

(iii) When the residue type is nucleobase and the first & the second binding locations are any one of sugar and phosphate, the atom is named as gamma.

Table 13 shows that among the 389 chelates, Mg2+ ions have a tendency to form bidentate chelates with maximum frequency.

Table 13: Distribution of Mg-binding sites in chelates

| One End of the Chilet | Other Ends of the Chilet | Denticity | No. of Sites |
|---|---|---|---|
| Nucleic Acid Base | Nucleic Acid Base | Bidentate | 115 |
| Nucleic Acid Base | Sugar/Phosphate | Bidentate | 32 |
| Sugar/Phosphate | Sugar/Phosphate | Bidentate | 215 |
| Nucleic Acid Base | Nucleic Acid Base | Tridentate | 4 |
| Nucleic Acid Base | Sugar/Phosphate | Tridentate | 2 |
| Sugar/Phosphate | Sugar/Phosphate | Tridentate | 21 |

# IV. Conclusion

Although interaction between the Mg2+ ions and RNA is quite well studied, it does not give complete understanding of the role of Mg2+ ions in RNA structure. For better understanding of the Mg-RNA complexes, in the present study, we have investigated several crystal structures of RNA for having Mg-RNA complexes. We have also generated reports for various statistics related to these Mg-RNA complexes by computing base pairs. Moreover, we have reported the capability of the Mg2+ ions in altering the stability of base pairs. For this, we have computed the e-values of the base pairs. Throughout our investigation, we have computed and reported the following statistics related to Mg-RNA complexes:

(a) General statistics regarding Mg-binding of nucleic acids related to bases in paired and unpaired states.

(b) General statistics related to interaction of Mg2+ ions with Watson-Crick base pairs & non-Watson-Crick base pairs
(c) The distance between the Mg2+ ion and the atom of a base pair with which the Mg2+ ion is bound.
(d) Statistics regarding the interaction of Mg2+ ions with base/sugar/phosphate and the atoms which are involved in such interaction.
(e) Statistics regarding the position of a base pair in the secondary structure of RNA like helix, coil, loop etc. with which Mg2+ ion has been interacted.
(f) Statistics for binding of Mg2+ ions with the atoms of any one edge among the three edges (Watson-Crick, Hoogsteen and Sugar) of a base.
(g) Statistics related to formation of chelates of nucleic acid with the Mg2+ ions.

Still there are lots of metal ions like Na+, K+, Ca2+, Mn2+, Zn2+ etc. which have contributions to the stability of RNA structures; we have not analyzed them yet. Our limitation is due to unavailability of sufficient data of RNA structures with these ions. Strength of interaction between the ions and RNA base pair could be also understood by performing intensive quantum chemical calculations by DFT or yet advanced methods. It would be interesting to address such studies also in future. We have only analyzed data that have been taken from the PDB database.

We have only analyzed base pairs but base triplets, base quartates etc. can also be analyzed when more such data is available .

## Funding Information:

This research is not supported by any kind of funding.

## References

1. Vernon W.B. (1988) The role of magnesium in nucleic-acid and protein metabolism. Magnesium 7(5-6):234-248.
2. Every A.E., Russu I.M. (2008) Influence of magnesium ions on spontaneous opening of DNA base pairs. Journal of Physical Chemistry B 112(25):7689-7695.
3. Misra Vinod K., Draper David E. (1998) On the role of magnesium ions in RNA stability. Biopolymers (Nucleic Acid Sciences) 48(2-3):113–135.
4. Mukherjee S., Bhattacharyya D. (2013) Influence of divalent magnesium ion on DNA: molecular dynamics simulation studies. Journal of Biomolecular Structure and Dynamics 31(8):896–912.
5. Pasternak K., Kocot J., Horecka A. (2010) Biochemistry of magnesium. Journal of Elementology 15(3):601-616.
6. Yamagami R., Sieg J.P., Bevilacqua P.C. (2021) Functional Roles of Chelated Magnesium Ions in RNA Folding and Function. Biochemistry 60(31):2374-2386.
7. Halder A., Roy R., Bhattacharyya D., Mitra A. (2017). How Does $Mg^{2+}$ Modulate the RNA Folding Mechanism: A Case Study of the G:C W:W Trans Basepair. Biophysical Journal 113:1-13.
8. Halder A., Roy R., Bhattacharyya D., Mitra A. (2018). Consequences of Mg2+ binding on the geometry and stability of RNA base pairs. Physical Chemistry Chemical Physics 20:1-15.
9. Zagórski W., Filipowicz W., Wodnar A., Leonowicz A., Zagórska L., Szafrański P. (1972) The Effect of Magnesium-Ion Concentration on the Translation of Phage-f2 RNA in a Cell-Free System of Escherichia coli. The FEBS Journal 25(2): 315-322.
10. Sosunov V., Sosunova E., Mustaev A., Bass I., Nikiforov V., Goldfarb A. (2003) Unified two-metal mechanism of RNA synthesis and degradation by RNA polymerase. EMBO J. 22(9):2234-2244.
11. Leontis N. B., Westhof E. (2001) “Geometric nomenclature and classification of RNA base pairs.” *RNA, Cambridge University Press* 7(4):499-512.
12. Zheng H., Shabalin I.G., Handing K.B., Bujnicki J.M., Minor W. (2015) Magnesium-binding architectures in RNA crystal structures: validation, binding preferences, classification and motif detection. Nucleic Acids Research 43(7):3789–3801.
13. Auffinger P., Grover N., Westhof E. (2011) Metal Ion Binding to RNA. Metal Ions in Life Sciences 9:1-35.
14. Robinson H., Gao Y.-G., Sanishvili1 R., Joachimiak A., Wang A. H.-J. (2000) Hexahydrated magnesium ions bind in the deep major groove and at the outer mouth of A-form nucleic acid duplexes. Nucleic Acids Research 28(8):1760-1766.

15. Batey R.T., Doudna J.A. (2002) Structural and energetic analysis of metal ions essential to SRP signal recognition domain assembly. Biochemistry 41:11703-11710.
16. Schrodinger, L., and DeLano W. (2020) "PyMOL." Available at: http://www.pymol.org/pymol.
17. Petrov A.S., Lamm Gene, Pack G.R. (2002) Water-Mediated Magnesium-Guanine Interactions. The Journal of Physical Chemistry B 106(12):3294–3300.
18. Leontis NB, Zirbel CL (2012) Nonredundant 3D structure data-sets for RNA Knowledge extraction and benchmarking. RNA 3D structure analysis and prediction. Springer, pp 281–298. https://doi.org/10.1007/978-3-642-25740-7_13)
19. Roy P., Bhattacharyya D. (2022) "MetBP: a software tool for detection of interaction between metal ion–RNA base pairs." *Bioinformatics* 38(15):3833-3834, https://doi.org/10.1093/bioinformatics/btac392
20. Das J., Mukherjee S., Mitra A., Bhattacharyya D. (2006) Non-canonical Base Pairs and Higher Order Structures in Nucleic Acids: Crystal Structure Database Analysis. J. Biomol. Struct. Dynam. 24(2):149-161.
21. Panigrahi S., Pal R., Bhattacharyya D. (2011) Structure and Energy of Non-Canonical Basepairs: Comparison of Various Computational Chemistry Methods with Crystallographic Ensembles, Journal of Biomolecular Structure and Dynamics 2(3):541-556.
22. Das S., Roy S., Bhattacharyya D. (2022) Understanding the role of non-Watson-Crick base pairs in DNA–protein recognition: Structural and energetic aspects using crystallographic database analysis and quantum chemical calculation, Biopolymers 113(7):e23492.